# ANALYSIS OF SYNCHRONIZATION MECHANISMS IN OPERATING SYSTEMS

Oluwatoyin Kode[1] and Temitope Oyemade[2]

Department of Computer Science, Bowie State University, Maryland, USA

[1]okode@bowiestate.edu
[2]toyemade@bowiestate.edu

## ABSTRACT

*This research analyzed the performance and consistency of four synchronization mechanisms—reentrant locks, semaphores, synchronized methods, and synchronized blocks—across three operating systems: macOS, Windows, and Linux. Synchronization ensures that concurrent processes or threads access shared resources safely, and efficient synchronization is vital for maintaining system performance and reliability. The study aimed to identify the synchronization mechanism that balances efficiency, measured by execution time, and consistency, assessed by variance and standard deviation, across platforms. The initial hypothesis proposed that mutex-based mechanisms, specifically synchronized methods and blocks, would be the most efficient due to their simplicity. However, empirical results showed that reentrant locks had the lowest average execution time (14.67ms), making them the most efficient mechanism, but with the highest variability (standard deviation of 1.15). In contrast, synchronized methods, blocks, and semaphores exhibited higher average execution times (16.33ms for methods, 16.67ms for blocks), but with greater consistency (variance of 0.33). The findings indicated that while reentrant locks were faster, they were more platform-dependent, whereas mutex-based mechanisms provided more predictable performance across all operating systems. The use of virtual machines for Windows and Linux was a limitation, potentially affecting the results. Future research should include native testing and explore additional synchronization mechanisms and higher concurrency levels. These insights help developers and system designers optimize synchronization strategies for either performance or stability, depending on the application's requirements.*

## 1. INTRODUCTION

The study of synchronization mechanisms has become increasingly critical with the growing complexity of modern operating systems. Synchronization ensures that concurrent processes or threads access shared resources in a controlled and orderly manner, preventing issues such as race conditions and data corruption [1][3]. In multi-threaded or multi-process environments, efficient synchronization tools are essential for maintaining system performance, scalability, and reliability. However, developing efficient and reliable synchronization mechanisms involves several challenges, such as competing for shared resources and locks, and the overhead associated with synchronization. Understanding the performance characteristics, scalability limitations, and optimization opportunities of synchronization tools is essential [2] [4].

Mechanisms such as reentrant locks, semaphores, and synchronized blocks are widely used in modern operating systems, including macOS, Windows, and Linux.

### 1.1. Research Objectives

The primary objective of this research is to conduct a comprehensive analysis of synchronization mechanisms in modern operating systems. The study focuses on four widely used mechanisms, reentrant locks, semaphores, synchronized methods, and synchronized blocks. The research aims to identify which mechanisms offer the best balance between performance and consistency by evaluating their performance across three major operating systems, macOS, Windows, and Linux (Ubuntu 22.02). Specific objectives include:

**1.1.1.** Reviewing and evaluating existing synchronization mechanisms and tools implemented in modern operating systems.

**1.1.2.** Assessing the performance of the synchronization mechanisms by calculating the average execution times of each synchronization mechanism under different system configurations.

**1.1.3.** Analyzing the variability in performance across different platforms and Identifying the most efficient mechanism for high-concurrency environments and the most stable mechanism for systems prioritizing consistency.

## 1.2. Research Hypothesis

Based on the popularity of mutex forms of synchronization, it is hypothesized that synchronized block and synchronized method are likely to be the most efficient mechanisms, followed by semaphore and reentrant lock.

## 1.3. Structure of the Research Paper

The structure of this paper includes several key sections. The introduction provides context and objectives for the research. A literature review follows, which examines existing studies on synchronization mechanisms and highlights gaps in the research. The methodology section details the experimental setup, explaining the implementation of each synchronization mechanism in Java, the platforms used, and the data collection process. The results and analysis section presents the empirical findings, including mean execution times, standard deviation, and variance for each synchronization mechanism, along with an analysis of their performance across different operating systems. The discussion interprets these results, with particular emphasis on their implications for developers and system designers, addressing the trade-offs between performance and consistency. The paper concludes with a summary of the key findings and suggestions for future work.

## 1.4. Limitations of the Research

This study has some limitations. First, the use of virtual machines to run both Windows and Linux may introduce performance overheads that do not occur in native environments. This could affect the accuracy of comparisons with macOS, which was tested in its native environment. Additionally, the sample size is limited to four synchronization mechanisms, excluding other potential methods such as lock-free data structures or software transactional memory. Lastly, platform-specific optimizations, such as differences in thread scheduling and memory management across operating systems, may influence the results. This study assumes general-purpose optimization for all operating systems tested, but in specialized environments, performance could vary significantly.

Despite these limitations, the research provides valuable insights into the efficiency and variability of synchronization mechanisms across different operating systems. It offers a foundation for further studies and optimizations in concurrent computing environments, helping developers and system designers make informed decisions about the synchronization methods that best suit their applications.

## 2. LITERATURE REVIEW

Previous studies have explored synchronization mechanisms such as mutual exclusion (mutex) locks, semaphores, and barriers. Various synchronization algorithms, including Peterson's algorithm and Dekker's algorithm, have been proposed to address concurrency challenges. However, the performance of synchronization tools in real-world operating systems remains an area of active research. This paper comprehensively reviews various studies conducted by different authors to pinpoint prevalent synchronization mechanisms.

Singh et al. [5], in their study of comparing the performance of various Software Transactional Memory (STM) concurrency control protocols, evaluated the performance of Basic Timestamp Ordering (BTO), Serialization Graph Testing (SGT), and Multi-Version Time-Stamp Ordering (MVTO) concurrency control protocols within an STM library. The authors also evaluated these protocols in comparison with the linked-list module from the Synchrobench benchmark, which encompasses lock-coupling list, lazy-list, ESTM (Elastic Software Transactional Memory), and lock-free list. A transactional SET data structure was implemented to test the STM library. The results showed that BTO and MVTO generally outperformed SGT and other Synchrobench implementations in terms of CPU time per thread.

In another study by Souto and Castro [7], they discussed operating systems for lightweight manycore processors designed for high parallelism and energy efficiency. They proposed reshaping OS-level abstractions to reduce memory footprint without significant overhead, introducing a cooperative time-sharing lightweight task model. They also presented a task-based execution engine that supported numerous OS-level execution flows with reduced memory consumption. Additionally, the study included an evaluation of the proposed engine, showing advantages in memory usage, core utilization, and performance on real-world applications. The article presented an in-depth analysis of enhancing concurrency and optimizing memory usage in distributed operating systems designed for lightweight manycore architectures.

Yastrebenetsky and Trakhtenbrot [6] assessed the Synchronization Complexity Metric (SCM) against Weyuker's soundness properties and Zuse's software measurement scales, demonstrating its utility in evaluating complex software systems. SCM helped determine the number of tests required for proper coverage in concurrent program testing and facilitated comparison between system implementations based on synchronization complexity. The paper distinguished between intentional and unintentional interleaving, emphasizing the importance of deliberate interleaving caused by explicit synchronization statements in code. The metric's growth was analyzed through real-world applications, indicating that SCM did not increase exponentially and was an essential tool for testers and software developers to estimate testing efforts and improve system design.

Kim et al. [8] analyzed Read-Copy-Update (RCU) style non-blocking synchronization mechanisms on manycore-based operating systems, focusing on performance and scalability. They compared the traditional RCU with newer mechanisms like Read-Log-Update (RLU) and Multi-Version Read-Log-Update (MV-RLU), highlighting their APIs, operation procedures, and performance in various scenarios. The study conducted extensive performance evaluations using micro-benchmarks on the sv6 research operating system, revealing that MV-RLU generally offered better performance. The paper discussed the potential for adopting RCU-style mechanisms in operating systems and noted the importance of careful implementation to avoid performance bottlenecks. The paper provided valuable insights into the effectiveness of different synchronization mechanisms in manycore environments.

Inglés et al. [9] discussed a C++ framework for high-performance data acquisition systems, emphasizing the use of multithreading and ring-buffer data structures for concurrent data transfer. They evaluated various C++ frameworks based on latency measurements, focusing on the time interval between data publication and gathering in a one-producer-two-consumer setup. The study compared the performance of different frameworks, including CxxDisruptor and FastFlow, and suggests that a simple framework using standard C++ libraries can achieve satisfactory results. The authors recommended optimizations such as using a Linux Real-Time Kernel and shared-memory technologies to improve message transfer reliability and avoid data loss. The paper concludes that the efficient use of multithreading in C++ can meet the high accuracy and performance demands of real-time data acquisition systems.

Alagalla et al. [3] discussed the importance of synchronization in Distributed Real-Time Operating Systems (RTOS) and the challenges in enhancing their efficiency and accuracy. The authors reviewed various synchronization techniques, including lock-based, semaphore-based, and hardware-oriented methods, and identified the problems faced by current Distributed RTOS. They also proposed multiple strategies to improve synchronization efficiency, focusing on multi-core processors and Internet of Things (IoT) based networks. The paper suggested further research to test and validate the proposed techniques, with an emphasis on kernel-level system development for IoT and embedded systems.

Joseph and Dhanwada [10] proposed a unique process synchronization approach that did not rely on atomic operations or interrupt disabling. Their study demonstrated the effective use of non-caching, shared on-chip memory to reduce synchronization overheads as the number of processor cores increased. The study validated the proposed mechanism through experiments on a virtual prototyping platform, showing its effectiveness compared to external memory-based schemes. The study also suggested potential extensions for the signaling scheme to support multiple signals and inter-core communication types.

Park et al. [11] discussed the importance of kernel synchronization primitives, such as kernel locks, for OS design, performance, and correctness. They introduced the concept of application-informed kernel synchronization primitives, which allowed application developers to influence kernel behavior. The paper presented SynCord, a framework that enabled the modification of kernel locks without recompiling or rebooting, providing APIs for user-defined lock policies. SynCord was evaluated to show minimal runtime overhead and performance comparable to state-of-the-art locks, significantly enhancing application performance.

Chehab et al. [12] introduced the Compositional Lock Framework (CLoF), a framework designed to support the entire memory hierarchy in multi-level Non-Uniform Memory Access (NUMA) systems, addressing the challenges posed by different NUMA architectures. CLoF aimed to optimize locking mechanisms for many-core servers, exploiting the performance promises of such systems while ensuring correctness, especially on Weak Memory Models (WMM) like Armv8. The paper provided an inductive argument, verified with model checkers, to demonstrate the correctness of CLoF on WMMs. It also included an evaluation showing that CLoF locks outperformed existing NUMA-aware locks in most scenarios. CLoF leveraged composability to generate numerous correct-by-construction NUMA-aware locks and supported heterogeneity by allowing different spinlock implementations across the memory hierarchy levels. The paper presented a significant contribution to concurrent programming and synchronization for modern server architectures.

Felber et al. [13] presented Hardware Read-Write Lock Elision (RW-LE), a novel Hardware Lock Elision (HLE) technique for read-write locks, which leveraged POWER8's microarchitectural features to enhance parallelism in concurrent applications. RW-LE introduced a hardware-software co-design that utilized suspend/resume and rollback-only transactions to optimize read-write lock elision. The technique offered significant performance improvements, with up to tenfold speedups over traditional HLE approaches, especially in read-intensive workloads. RW-LE allowed writers to execute concurrently with readers, using speculative buffering of memory writes and an RCU-like quiescence mechanism to ensure consistency. The paper included a thorough experimental study with benchmarks and real-life applications, demonstrating RW-LE's effectiveness across various scenarios.

In their article, Kurdi et al. [14] discussed service sharing in distributed systems, focusing on mutual exclusion to minimize wait times and guarantee fair access to Critical Sections (CS). They introduced a token-based approach and the Fairness Algorithm for Priority Process (FAPP) to manage prioritized processes and improve efficiency. The system architecture allowed clients to request CS access based on priority, which was determined by shopping cart contents. Evaluation experiments demonstrated the effectiveness of the FAPP approach in reducing average waiting times. Overall, the proposed system aimed to enhance service sharing by organizing access based on priority rather than solely on arrival time.

Ling Yang [15] explored the significance of mutual exclusion and condition synchronization in concurrent programming. The study discussed how these concepts were essential for achieving inter-thread communication and synchronization, especially in concurrent programs where multiple threads ran simultaneously. The research proposed countermeasures for classic synchronization problems like the producer/consumer and dining philosophers' problems, emphasizing the importance of mutual exclusion and condition synchronization. Through practical implementations and comparisons, the study demonstrated how these synchronization techniques were crucial in ensuring proper concurrency management. The authors recommended further research in concurrent programming, recommending exploration of areas such as the readers/writers problem, the sleeping barber problem, and the cigarette smokers problem.

Choi and Seo [16] introduced a novel mutex mechanism designed to address Inter Process Communication (IPC) issues in consumer electronics systems, specifically focusing on mitigating starvation of low-priority processes without sacrificing responsiveness for high-priority processes. The proposed mechanism, called eclectic_mutex, dynamically adjusted process priorities based on waiting time to balance fairness and responsiveness. Evaluation results demonstrated that eclectic_mutex effectively prioritized high-priority processes while limiting the waiting time for low-priority processes, offering a viable alternative to existing mutex schemes like default mutex and rt_mutex.

Alagalla and Rajapaksha [17] discussed the importance of synchronization in Distributed Real-Time Operating Systems (DRTOS) and proposed various techniques to enhance their efficiency and performance. They identified challenges faced by DRTOS and suggested solutions such as priority-based task synchronization, thread-level parallelism, consensus algorithms, and mutex handling. Additionally, they discussed the use of FreeRTOS for inter-process synchronization and hardware resource synchronization, such as semaphores for hybrid DRTOS. Overall, the research aimed to improve data processing performance in embedded systems, IoT, and distributed processing networks.

Sai et al. [18] discussed the utilization of thread pools to address the producer-consumer problem in scenarios where multiple short tasks needed to be processed concurrently. By employing semaphores to manage access to shared buffers, they enhanced the efficiency of the proposed model. The study outlined various scenarios where this approach could be applied, focusing mainly on real-life examples like delivery systems. Through the implementation of a thread-pool-based solution, the paper demonstrated improved efficiency compared to traditional thread-based approaches and discussed methods to determine the optimal size of the thread pool. Overall, the study provided insights into enhancing concurrency and efficiency in server programs dealing with repetitive operations on multiple short tasks.

Huang and Hwang [19] proposed a semaphore-based programming model and scheduling algorithm to address constrained parallelism in task graph programming. They addressed the limitations of existing Task Graph Programming Models (TGPMs), which often overlooked constrained parallelism within task graphs. The semaphore model allowed users to specify constraints, such as limiting the number of concurrent tasks or resolving conflicts between tasks using semaphores. The article outlined the motivation behind the research, driven by the need for efficient parallelism in high-performance computing, particularly in functions like Computer-Aided Design (CAD) algorithms for integrated circuits. Experimental results demonstrated significant performance improvements in real-world CAD applications compared to existing heuristics. Overall, the semaphore model provided a streamlined and adaptable solution that could seamlessly integrate into current TGPMs, offering potential applications beyond CAD.

Ji and Song [20] formally analyzed Peterson's solution, a classical algorithm for mutual exclusion in concurrent systems. Peterson's solution, proposed by G.L. Peterson in 1981, addressed the problem of ensuring mutual exclusion in critical sections without requiring additional hardware mechanisms. Despite its acceptance as a classic algorithm, the formal analysis of its correctness, particularly regarding safety and liveness properties, remained incomplete. The study highlighted the importance of considering execution order in concurrent programming for correctness. This work laid the groundwork for extending the analysis to more concurrent processes and exploring liveness properties in Peterson's solution.

Dalmia et al. [21] focused on enhancing the efficiency and scalability of synchronization primitives for Graphics Processing Units (GPUs). They addressed the limitations of existing GPU synchronization primitives, particularly in scenarios with heavy contention between threads accessing shared synchronization objects. The authors proposed more efficient designs for barriers and semaphores tailored to the unique processing model of GPUs. These designs, such as multi-level sense-reversing barriers and priority mechanisms for semaphores, aimed to improve performance and scalability while avoiding livelock issues. The article presented detailed analyses and benchmarks demonstrating the effectiveness of the proposed designs, showcasing significant performance improvements compared to state-of-the-art solutions. Overall, the work highlighted the importance of optimizing synchronization primitives for modern GPU applications and provided insights into designing efficient synchronization mechanisms that leverage the characteristics of GPU architectures.

Albinali et al. [22] evaluated various synchronization algorithms used in multi-threaded web servers, which were crucial for efficiently handling numerous concurrent requests in the age of internet proliferation. Through empirical experiments simulating file synchronization in distributed systems, they compared traditional algorithms (Mutex locks, Spinlocks, Semaphore) with modern algorithms (WebR2sync) based on metrics such as Synchronization Time, Failure Rate, and Scalability. The results showed that Semaphore excelled in synchronization time, WebR2sync in failure rate, and Mutex in scalability. The study aided in selecting the most suitable synchronization algorithm based on specific server requirements, contributing to optimized server performance.

Fiestas and Bustamante [23] explored the performance of different mutex and barrier algorithms in multicore computing. They highlighted the challenges of ensuring coherence in such systems and the importance of synchronization and coordination among threads. The study evaluated various mutex algorithms, including Compare & Swap, Test & Set, Test & Test & Set, and Ticket Lock. It also assessed barrier algorithms like the Centralized Barrier and Sense-Reverse Barrier. The experiments conducted on an entry-level server and personal computers revealed that TATAS with exponential backoff performed exceptionally well among mutex algorithms, outperforming even the POSIX pthread library implementation. Similarly, the Sense-Reverse Barrier demonstrated superior performance compared to the Centralized Barrier, and the use of exponential backoff had varying effects on different barrier algorithms. These findings suggested that careful consideration of mutex and barrier algorithms could significantly impact the performance and correctness of parallelized code in multicore systems.

Zhang et al. [24] discussed the challenges and performance trade-offs associated with using synchronized locks versus reentrant locks in multi-threaded Java applications. They highlighted the labor-intensive and error-prone nature of manual refactoring and proposed an automated refactoring framework to transform synchronized locks into reentrant locks. The framework leveraged program analysis techniques and bytecode manipulation to ensure the consistency and correctness of the transformation process. Evaluation on various benchmarks demonstrated the effectiveness and practicality of the proposed approach. Key contributions included a performance comparison, a detailed refactoring process, the implementation of the refactoring tool, and evaluation on Java applications. The article concluded by outlining the motivation, methodology, and results of the experimentation, emphasizing the performance implications of lock mechanisms in different scenarios.

Huang and Fu [25] conducted a comprehensive analysis of Java locking mechanisms, focusing on CPU usage, memory consumption, and implementation ease. They categorized various locks, including synchronized blocks, synchronized methods, and reentrant locks, and evaluated their performance metrics. Notably, they observed differences in CPU usage among different lock types, with spin locks exhibiting the highest overhead due to their continuous consumption of CPU resources. They also highlighted variations in memory consumption, particularly noting that fair locks tend to consume more memory than non-fair locks due to maintaining a queue for thread access. Furthermore, the study delved into the exception-handling mechanisms and package dependencies associated with different locks. It provided insights into the underlying method implementations of locks, shedding light on how each lock type operates at a fundamental level. They concluded the research by recommending future work, including the addition of metrics like code complexity and the creation of intelligent models to guide lock selection. Additionally, they propose mitigating the impact of JVM optimizations in experimental setups to better reflect differences between lock types.

## 3. METHODOLOGY

This research adopted a multifaceted approach combining empirical evaluation and theoretical analysis to achieve its objectives. Figure 1 shows a flowchart representation of the methodology used in this research.

### 3.1. Literature Review

The methodology began with a comprehensive literature review of existing research on synchronization mechanisms in operating systems. This involved identifying relevant academic journals, conference proceedings, textbooks, and research papers that discuss synchronization primitives, algorithms, and tools. The literature review offered a comprehensive insight into the various synchronization mechanisms and highlighted prior research findings.

### 3.2. Experimental Setup and Empirical Evaluation

Our study compared the performance of three major operating systems in terms of synchronization, focusing on Java as our implementation platform. We devised a shared resource represented by a Java object containing two variables: sharedCounter of type integer and sharedColl of type collections. To conduct our tests, we utilized randomly generated integers from the SecureRandom library within the java.security package.

We established a SynchronizationMechnism.java interface featuring an abstract method named run(), returning a SharedData object. This interface served as the basis for four synchronization mechanism implementations, each providing its implementation of the run() method. Leveraging inheritance and polymorphism, we ensured uniformity in implementation while accommodating differences in lock methodologies.

Each class incorporates an instance of SecureRandom as a class variable and overrides the run() method defined in the SynchronizationMechanism interface. Within the run() method, a new instance of SharedData is created, threads are initialized to execute concurrent tasks, while the main thread holds for the tasks to finish with the join() method.

The runnable class defined the task to be executed by each thread through a lambda expression. This task involved running a loop 1000 times, populating the SharedData object with the counter value, and randomly generating integers using the SecureRandom object. Finally, the method returned the sharedData object after all threads have completed execution.

To assess latency across operating systems, we implemented four synchronization mechanisms in Java: semaphores, reentrant locks, and two forms of mutex (synchronized block and synchronized method). Our experiments will executed the Java code on Windows 11, Apple macOS, and Ubuntu 22.02, chosen for their widespread use among current users.

### 3.2.1. Semaphores

In Java, semaphores are implemented through the Semaphore class. This class provides constructors to initialize semaphores with an initial number of permits and methods to acquire and release permits. The acquire() method is used by threads to request permits, and if available, they are granted access; otherwise, the thread blocks until permits become available. Conversely, the release() method is used to return permits to the semaphore, making it available to other threads. Semaphores in Java can be either binary or counting semaphores, depending on how they are initialized. They offer a powerful mechanism for coordinating access to shared resources in concurrent Java programs, aiding thread synchronization management and avoiding race conditions [25][26].

#### 3.2.1.1. Implementation and Configuration of Semaphores

Java Class: Semaphore from the java.util.concurrent package.

Initial Configuration: A binary semaphore (with one permit) was used for controlling access to the critical section, ensuring that only one thread could enter at a time.

Method Calls:

acquire(): Each thread called this method to acquire the semaphore before accessing the shared resource. If no permits were available, the thread was blocked until a permit was released.

release(): After finishing the operation, the thread called release() to return the permit, allowing other threads to proceed.

Thread Behavior: Threads were blocked if the semaphore was unavailable, providing mutual exclusion.

### 3.2.2. Reentrant locks

In Java, a reentrant lock is a synchronization mechanism to regulate access to shared resources among multiple threads or processes. Unlike traditional locks, it enables the same thread to acquire the lock multiple times, known as "reentrancy" or "recursive locking," without risking deadlock situations. Upon acquiring the lock, a thread gains exclusive access to the protected resource, allowing nested locking without blocking itself. This behavior is especially useful for scenarios involving recursive method calls or invocation of other methods requiring the same lock, promoting code modularity and flexibility. The ReentrantLock class in Java provides a reliable implementation of reentrant locks, ensuring secure acquisition and release of locks by threads [24].

### 3.2.2.1. Implementation and Configuration of Reentrant locks

Java Class: ReentrantLock from the java.util.concurrent.locks package.

Initial Configuration: A single reentrant lock was created for the shared resource. The lock allowed the same thread to acquire it multiple times (reentrancy).

Method Calls:

lock(): Each thread called lock() before entering the critical section.

unlock(): After completing the operation, the threads called unlock() to release the lock, allowing other threads to acquire the lock.

Thread Behavior: This configuration allowed a thread to re-acquire the lock as needed.

### 3.2.3. Synchronized block

In Java, a synchronized block is a segment of code marked with the synchronized keyword, guaranteeing exclusive execution by a single thread by acquiring the intrinsic lock associated with a specified object or class. When a thread enters a synchronized block, it obtains the lock linked with the specified object, preventing other threads from accessing synchronized blocks on the same object until the lock is released. This mechanism guarantees thread safety within the synchronized block, allowing for the orderly execution of critical code sections [25][26].

### 3.2.3.1. Implementation and Configuration of Synchronized block

Java Keyword: synchronized block in Java.

Initial Configuration: The critical section was enclosed within a synchronized block, which locked on a shared object.

Thread Behavior: Only one thread could execute the code within the synchronized block at a time. Other threads were blocked until the lock on the object was released.

### 3.2.4. Synchronized method

In Java, a synchronized method is declared with the synchronized keyword, allowing only one thread to execute it at a time by acquiring the intrinsic lock associated with the object on which the method is called. When a thread invokes a synchronized method, it obtains the lock linked with the object, blocking other threads from executing synchronized methods on the object till the lock has been released. This mechanism guarantees that synchronized methods of an object are executed exclusively by one thread at a time, ensuring thread safety [25][26].

### 3.2.4.1. Implementation and Configuration of Synchronized method

Java Keyword: synchronized keyword applied to the entire method.

Initial Configuration: The method handling the shared resource was declared as synchronized, which means that the intrinsic lock on the object was automatically acquired when a thread called the method.

Thread Behavior: Only one thread could execute the synchronized method at any given time. Other threads attempting to call the method would be blocked until the current thread completed execution and released the lock.

Synchronized block and Synchronized method use mutex algorithms.

### 3.2.5. Parameters and Conditions:

Number of Threads: 10 threads were used in each test to simulate concurrent access to shared resources.

Number of Iterations: Each thread executed the task 1000 times.

Random Number Generation: SecureRandom was used to generate random integers in each iteration, simulating a dynamic workload.

Operating Systems specification: The tests were conducted on the following platforms:

macOS (M1 chip with macOS Sonoma) with 16GB RAM, and 512GB SSD Storage.

Windows 11 (via Parallels virtual machine)

Linux Ubuntu 22.02 (via Parallels virtual machine)

The parallel virtual machines were set up on the same hardware as the macOS. The goal was to run the tests on a machine with the same hardware specification, although the virtual machine may not simulate a native environment 100%.

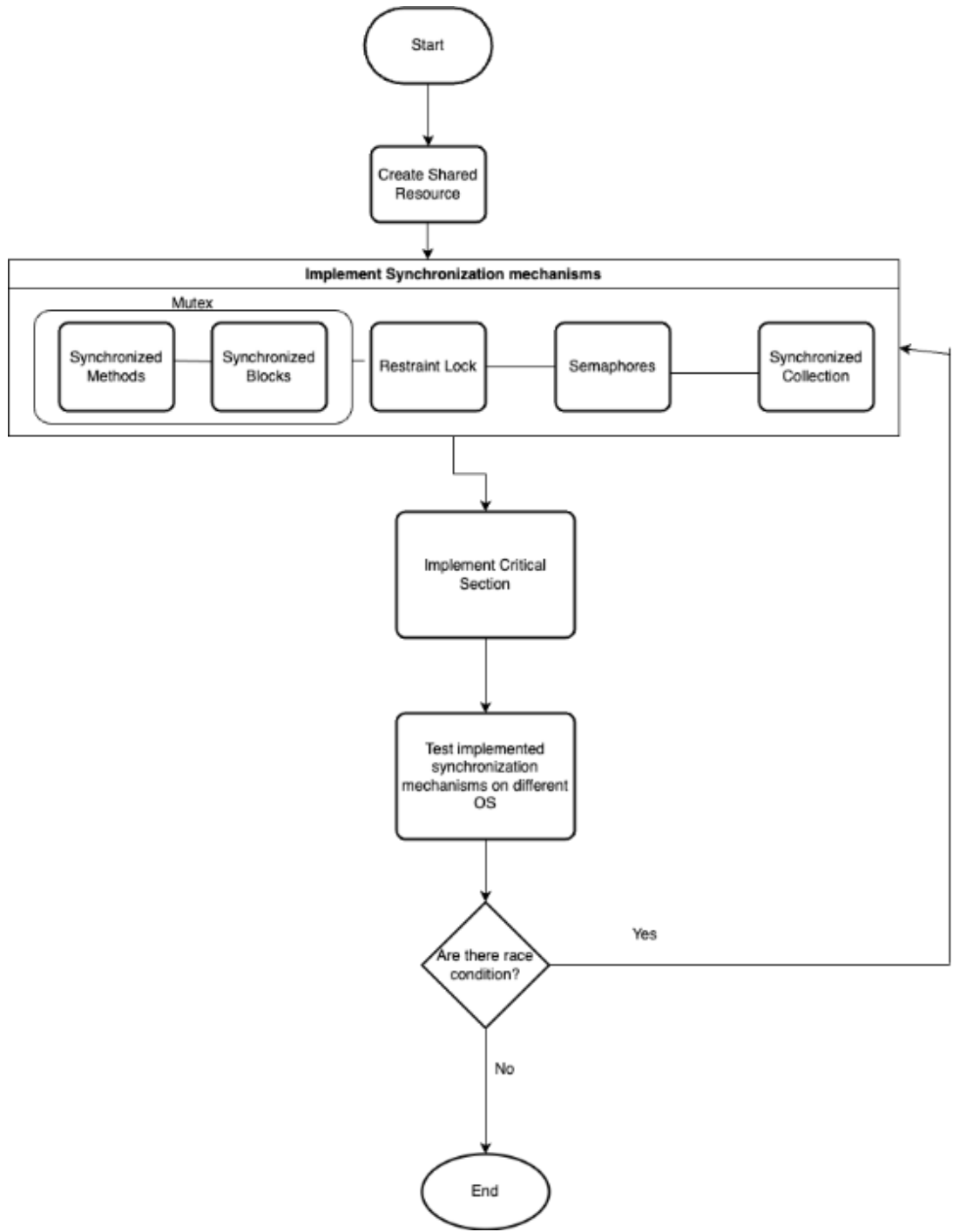

Figure 1. Flowchart of design methodology

### 3.3. Data Analysis and Interpretation

The collected performance data was analyzed to identify trends. Data analysis provided quantitative insights into the synchronization mechanisms' performance, scalability, and efficiency under different conditions. The empirical findings and theoretical insights were interpreted and synthesized to draw conclusions and implications for operating system design and optimization. The strengths, weaknesses, and trade-offs of synchronization tools were discussed, and recommendations for improving synchronization mechanisms were proposed based on the research findings.

We implemented a test class named SynchronizationMechnismTest.java, which contained all test cases for the different synchronization mechanisms. We used the Java secure random library to generate 1000 random integers for the data set. To ensure accuracy, we ran each test 5 times and recorded the time. The tests were also performed with 10 threads. Time measurements were taken by capturing the system time before and after each execution, and then calculating the difference in milliseconds. After recording the time for the 5 test runs, we calculated the average time using the formula $y = \frac{1}{n} \sum_{i=1}^{n} X_i$ where $y$ is the average time and $X_i$ is the time for each test iteration. The results are presented in the result section below.

## 4. RESULT

Upon conclusion of the research, an in-depth understanding of the design, implementation, and performance characteristics of synchronization tools in modern operating systems was gained. Factors influencing the performance and scalability of synchronization mechanisms were identified. The effectiveness of the synchronization mechanisms of the Linux, Windows, and macOS operating systems was also identified. The results of the tests are shown in Tables 1 to 3 and Figure 2 shows a graphical representation.

In Table 1, the time taken by the four synchronization mechanisms to execute on Apple Macintosh is displayed. The synchronized method took 16ms, the synchronized block took 17ms, the semaphore took 15ms, and the reentrant lock took 14ms.

In Table 2, the time taken by the four synchronization mechanisms to execute on Windows OS is displayed. The synchronized method took 16ms, the synchronized block took 17ms, the semaphore took 16ms, and the reentrant lock took 14ms.

In Table 3, the time taken by the four synchronization mechanisms to execute on Linux (Ubuntu 22.02) is displayed. The synchronized method took 17ms, the synchronized block took 16ms, the semaphore took 16ms, and the reentrant lock took 16ms.

| Synchronization Mechanism | Time Taken (in milliseconds) |
| --- | --- |
| Synchronized Method | 16 |
| Synchronized Block | 17 |
| Semaphore | 15 |
| Reentrant Lock | 14 |

Table 1. Result of the Synchronization Mechanism on Apple Macintosh

| Synchronization Mechanism | Time Taken (in milliseconds) |
|---|---|
| Synchronized Method | 16 |
| Synchronized Block | 17 |
| Semaphore | 16 |
| Reentrant Lock | 14 |

Table 2. Result of the Synchronization mechanism on Windows OS

| Synchronization Mechanism | Time Taken (in milliseconds) |
|---|---|
| Synchronized Method | 17 |
| Synchronized Block | 16 |
| Semaphore | 16 |
| Reentrant Lock | 16 |

Table 3. Result of the Synchronization mechanism on Linux (Ubuntu 22.02)

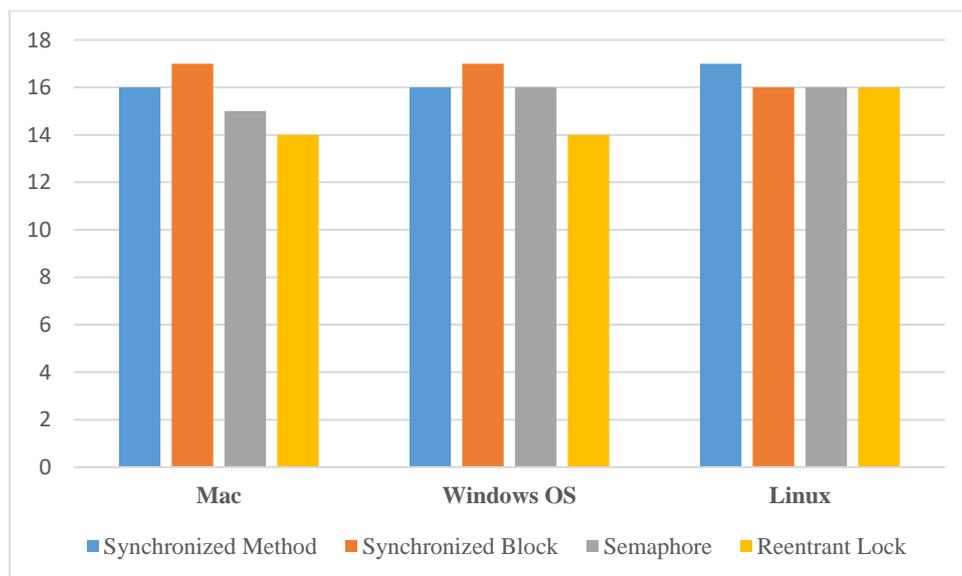

Figure 2. Graphical representation of the performance of the four synchronization mechanisms on Mac, Windows, and Linux operating systems

## 4.1. Statistical Analysis of Test Result

The mean, standard deviation, and variance of the synchronization mechanism results were analyzed to determine whether the variations hold any statistical significance. The results are depicted in Table 4 below. The synchronized method shows a mean, standard deviation, and variance of 16.33, 0.58, and 0.33, respectively. Synchronized block shows a mean, standard deviation, and variance of 16.67, 0.58, and 0.33 respectively. Semaphore shows a mean, standard deviation, and variance of 15.67, 0.58, and 0.33, respectively, and re-entrant lock shows a mean, standard deviation, and variance of 14.67, 1.15, and 1.33, respectively.

| Synchronization Mechanism | Mean | Standard Deviation | Variance |
|---|---|---|---|
| Synchronized Method | 16.33 | 0.58 | 0.33 |
| Synchronized Block | 16.67 | 0.58 | 0.33 |
| Semaphore | 15.67 | 0.58 | 0.33 |
| Reentrant Lock | 14.67 | 1.15 | 1.33 |

Table 4. Result of the statistical analysis of the synchronization mechanism

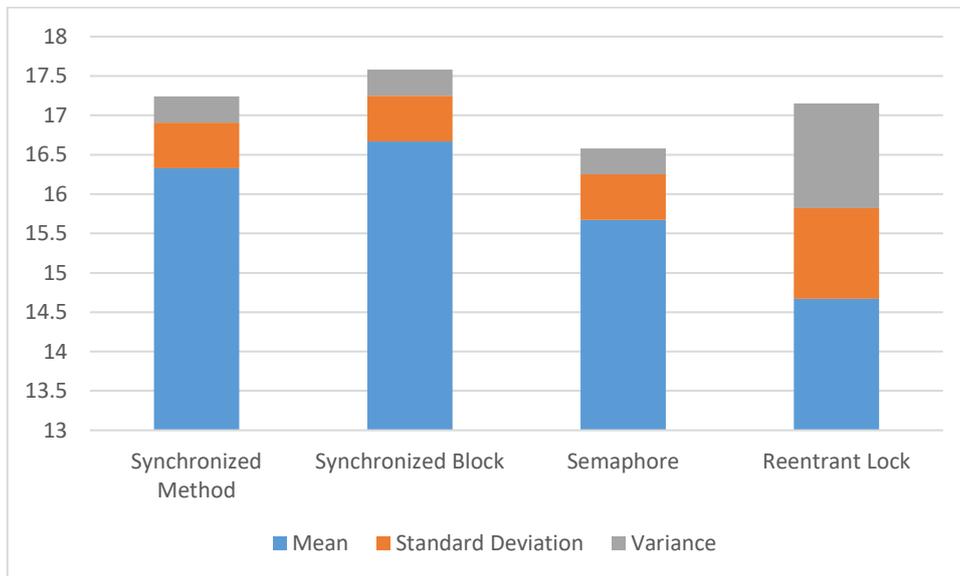

Figure 2. Graphical representation statistical analysis of the synchronization mechanism

# 5. DISCUSSION OF RESULT

The performance analysis of the four synchronization mechanisms, reentrant locks, semaphores, synchronized methods, and synchronized blocks revealed distinct differences in execution times and consistency across the three operating systems: macOS, Windows, and Linux. These differences were evaluated through raw scores and statistical analysis, which included calculating means, standard deviations, and variances.

The reentrant lock showed the lowest average execution time across the tested platforms, with a mean of 14.67ms. This suggests that reentrant locks are highly efficient at handling thread synchronization, particularly in high-concurrency environments where speed is a critical factor. However, it also exhibited the highest variance (1.33) and standard deviation (1.15), indicating greater variability in performance across operating systems. The variance in performance is likely due to platform-specific differences in how macOS, Windows, and Linux manage thread scheduling and memory. For example, macOS may have more optimized support for reentrant locks, allowing it to handle thread-switching and locking mechanisms more efficiently than Linux. This variability means that while reentrant locks offer better raw performance, they are less predictable across platforms, which may pose challenges in environments where consistent performance is required.

In contrast, synchronized methods and synchronized blocks displayed higher average execution times, with means of 16.33ms and 16.67ms, respectively. These mechanisms had lower standard deviations (0.58) and variance (0.33), indicating that their performance was more consistent across the three operating systems. This consistency can be attributed to their simpler locking mechanisms, which rely on intrinsic locks within objects or methods. Although they are slower, synchronized methods and blocks offer stable, predictable behavior, which is crucial for applications where maintaining uniform performance across platforms is a priority. The small variance suggests that these mechanisms handle thread contention and shared resource access in a relatively uniform manner, regardless of the operating system's underlying architecture.

Semaphores, with a mean execution time of 15.67 ms, performed in between reentrant locks and synchronized methods/blocks. Like the mutex-based mechanisms, semaphores also exhibited low variance (0.33) and standard deviation (0.58), making them a consistent choice across platforms. Semaphores are often used to control access to shared resources by limiting the number of threads that can access a critical section at once. This mechanism likely benefits from the fact that it can handle concurrency more flexibly than strict mutual exclusions, resulting in relatively low overhead while maintaining stability. Given their balance between performance and consistency, semaphores may be ideal for applications that need moderate concurrency with stable performance across different operating systems.

## 5.1. Platform-Specific Performance

The higher variance observed in the reentrant lock's performance may be due to platform-specific optimizations or inefficiencies in handling thread scheduling, context switching, and locking mechanisms. On systems like macOS, which are known for efficient thread management, reentrant locks performed exceptionally well, potentially benefiting from optimized locking algorithms and better support for concurrent operations. In contrast, on Linux, which prioritizes versatility and customizability, the performance of reentrant locks may suffer from higher overhead associated with managing multiple threads, leading to greater variability in execution times.

On the other hand, the consistent performance of synchronized methods, blocks, and semaphores across all platforms suggests that these mechanisms are less sensitive to operating system-specific optimizations. This is likely because their simpler locking mechanisms rely more on basic OS-level synchronization primitives, which are uniformly implemented across platforms. Therefore, they avoid the platform-specific bottlenecks or optimizations that affect reentrant locks.

## 5.2. Implications for System Designers and Developers

The findings of this research hold important implications for system designers and developers, particularly in choosing the appropriate synchronization mechanism based on the needs of their applications and target platforms.

### 5.2.1. Reentrant Locks for Critical Applications

For applications where raw performance is critical and variability in execution time is acceptable, reentrant locks offer the best performance. However, developers should be aware of the higher variance in performance across different platforms. If the application is being developed for a specific operating system, particularly one like macOS where reentrant locks perform well, this may be an ideal choice. However, if cross-platform consistency is a requirement, reentrant locks may not be suitable without additional optimization.

### 5.2.2. Synchronized Methods, Blocks, and Semaphores for Stability

For applications that prioritize stability and consistency across different operating systems, synchronized methods, synchronized blocks, and semaphores are the better options. Their low variance and standard deviation make them ideal for cross-platform applications where maintaining predictable performance is more important than achieving the lowest possible execution time. This is especially important in enterprise-level applications, where uniformity and reliability are critical, such as database systems or distributed computing environments.

### 5.2.3. Platform-Specific Optimizations

The differences in synchronization mechanism performance across macOS, Windows, and Linux highlight the importance of platform-specific optimizations. Developers working on cross-platform applications should be mindful of how each operating system handles thread management and synchronization. Performance tuning for each target platform may be necessary to achieve optimal results. For instance, Linux's performance issues with reentrant locks could potentially be mitigated by kernel-level optimizations or custom threading libraries.

## 6. CONCLUSION

The results of this study challenge the initial hypothesis that mutex-based synchronization mechanisms, such as synchronized blocks and synchronized methods, would be the most efficient forms of synchronization. It was predicted that these mechanisms, given their widespread use and simplicity, would outperform other methods such as semaphores and reentrant locks. However, the empirical findings showed that reentrant locks, rather than synchronized blocks or methods, consistently had the lowest average execution times, making them the most efficient in terms of raw performance.

The contradiction in the hypothesis arises from an overestimation of the efficiency of mutex-based mechanisms. While synchronized blocks and methods are straightforward and widely used for their simplicity and reliability, they introduce overhead through object-level locking and stricter mutual exclusion controls. These characteristics make them more stable but less efficient in high-concurrency, performance-critical environments. On the other hand, reentrant locks, despite being more complex, offer greater flexibility and better performance in these environments, particularly when the same thread needs to lock and unlock a resource multiple times. The higher variance observed in reentrant lock performance suggests that their efficiency is more platform-dependent, which was not fully accounted for in the initial hypothesis.

The study also revealed that semaphores, while offering a balance between mutexes and reentrant locks, did not outperform the reentrant lock, as the hypothesis had suggested. Although semaphores were expected to have a lower performance than mutexes, their ability to manage multiple threads accessing shared resources efficiently placed them closer to reentrant locks in performance. This again demonstrates the need to consider the specific characteristics and requirements of different synchronization mechanisms when making predictions about their performance.

### 6.1. Potential Areas for Future Work and Improvement

There are several areas where future research could build upon or improve the findings of this study:

### 6.1.1. Native Testing Across Platforms

This study was limited by the use of virtual machines for Windows and Linux, which could have introduced performance overheads that do not exist in native environments. Future work should include testing each operating system in its native environment to eliminate the influence of virtualization. This would provide a more accurate comparison of synchronization mechanisms across platforms.

### 6.1.2. Expanding the Range of Synchronization Mechanisms

This study focused on four commonly used synchronization mechanisms (reentrant locks, semaphores, synchronized methods, and synchronized blocks). Future research could explore additional mechanisms, such as lock-free data structures, read-copy-update (RCU), or software transactional memory (STM). These alternative mechanisms may offer better performance or scalability under certain conditions, and their inclusion would provide a more comprehensive understanding of synchronization performance.

### 6.1.3. Exploring Platform-Specific Optimizations

The variability in reentrant lock performance across operating systems suggests that further research could explore platform-specific optimizations for each synchronization mechanism. For instance, tuning Linux's thread scheduling or investigating macOS's optimized support for reentrant locks could yield valuable insights into how these mechanisms can be further improved or adapted for specific operating systems.

### 6.1.3. Testing Under Higher Concurrency Loads

This study used 10 threads in each test, but future research could increase the scale of concurrency to better understand how these synchronization mechanisms perform under extreme loads. This would be particularly valuable in real-world applications such as databases or large-scale web servers, where thousands of threads may be competing for resources simultaneously.

In conclusion, while the hypothesis predicted that mutex-based synchronization mechanisms would be the most efficient, the study found that reentrant locks offered superior performance. This finding highlights the importance of considering both the specific characteristics of synchronization mechanisms and the platforms on which they are deployed. Future research should expand the scope of synchronization methods, explore platform optimizations, and test mechanisms under more extreme conditions to further improve our understanding of synchronization performance.